# Detection of a diffusive cloak via second-order statistics

**Milan Koirala** and **Alexey Yamilov**

*Physics Department, Missouri University of Science and Technology*

*Corresponding author: mk3wd@mst.edu*

We propose a scheme to detect the diffusive cloak proposed by Schittny et al [Science 345, 427 (2014)]. We exploit the fact that diffusion of light is an approximation that disregards wave interference. The long-range contribution to intensity correlation is sensitive to locations of paths crossings and the interference inside the medium, allowing one to detect the size and position, including the depth, of the diffusive cloak. Our results also suggest that it is possible to separately manipulate the first- and the second-order statistics of wave propagation in turbid media. © 2016 Optical Society of America

A cloak conceals an object by molding the flow of light in the surrounding volume [1]. Transformation optics [2, 3] has enabled one to design artificial materials with spatially varying permittivity and permeability to achieve cloaking effect, see Fig. 1(a). We will refer to this scheme as a ballistic cloak. A different kind of cloak has been proposed by Schittny et al [4, 5] – a reflecting object is hidden inside a turbid medium where the light propagates diffusively, i.e. via a random walk, Fig. 1b. The cloaking effect is achieved by properly increasing diffusivity of the shell $D'$. Hiding an object inside a scattering medium is usually not difficult – it suffices to bury an object at the depth of several transport mean free paths. Meanwhile, the opposite task of detecting an inhomogeneity, such as cancer, is the holy grail of biomedical optics [6]. Notwithstanding, it has been pointed out [4] that the cloaking effect can be achieved for a shallow object and with a thin cloak made of a statistically uniform material. Furthermore, the cloak is omni-directional and broadband. This behavior can be understood by realizing that the light intensity in turbid media is governed by the diffusion equation, which is much simpler than the Maxwell equations for the ballistic cloak. The Laplace equation describing diffusion is formally equivalent to that of the electric/magnetic potential in the electro-/magneto- statics. In the latter context, core-shell structures have been known to preserve the uniformity of the electric [7]/magnetic [8] field lines in both two (cylinder) and three (sphere) dimensions. For light diffusion, it suggests that the proper choice of the diffusion constant ensures a uniform intensity at the output surface of a slab containing the cloak. This result has been verified [4] and extended to heat cloaking [9, 10]. Recently, further study revealed that the diffusive cloak can be compromised under pulsed illumination [11].

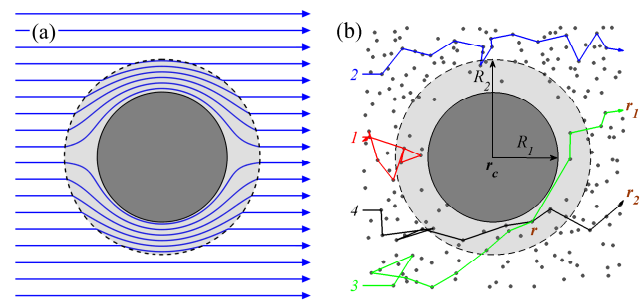

**Fig. 1.** (a) A schematic depiction of light paths in the ballistic cloak [2, 3]. Gray shell depicts the area where material properties have been modified based on principles of the transformation optics to mold the flow of light around the obstacle. (b) In turbid media, photons propagate in a random-walk-like fashion, with some being reflected (trajectory 1) and some – transmitted (trajectory 2). The diffusive cloak compensates the longer paths around the object by increased diffusivity (e.g. lower scatterer density) in the surrounding shell. In this work, we demonstrate that an interference due to trajectories that intersect throughout the volume (such as 3,4) can be exploited to detect the size and position of the embedded object.

The cloak in Fig. 1b is based on the diffusive description of wave transport. This is an approximation, which does not take into account wave interference [12, 13]. Second-order statistical properties such as fluctuations and correlations of intensity, contain information about the wave nature of transport, including the location where wave interference took place. We consider spatial intensity correlations defined as $C(\mathbf{r}_1, \mathbf{r}_2) = \langle \delta I(\mathbf{r}_1) \delta I(\mathbf{r}_2) \rangle / \langle I(\mathbf{r}_1) \rangle \langle I(\mathbf{r}_2) \rangle$, where $\delta I(\mathbf{r}) = I(\mathbf{r}) - \langle I(\mathbf{r}) \rangle$ is the deviation from the mean at point $\mathbf{r}$. Here, $\langle ... \rangle$ denotes the ensemble average that is accomplished by e.g. averaging over the wavelength of the source. The correlation is dominated by a short-range (on the order of transport mean free path, $\ell$) contribution $C_1(\mathbf{r}_1, \mathbf{r}_2)$ responsible for speckles [15]. We present a method for detecting the size and location of an object concealed by the diffusive cloak based on analysis of the long-range component of the correlation [16–18, 20] $C_2(\mathbf{r}_1, \mathbf{r}_2) \simeq C(\mathbf{r}_1, \mathbf{r}_2) - C_1(\mathbf{r}_1, \mathbf{r}_2)$ measured at the output surface of a slab of turbid medium.

We consider a coherent plane-wave illumination of a slab of lossless $2D$ turbid medium of thickness $L$ and transport mean free path $\ell$. Our goal is to compare the intensity and the long-range correlation in the slab (the reference) to those in a slab with an embedded reflecting cylinder (case I); and in a slab with the cylinder surrounded by the diffusive cloak (case II).

The ensemble-averaged intensity of the diffuse light inside the slab $\langle I(\mathbf{r}) \rangle$ satisfies the quasi-static diffusion equation [12]

$$D\nabla^2 \langle I(\mathbf{r}) \rangle = 0 \tag{1}$$

with boundary conditions $\langle I(y, 0) \rangle = I_0$ and $\langle I(y, L) \rangle = 0$. $D$ is the diffusion coefficient and $\mathbf{r} = (y, z)$ where $y$ and $z$ are the transverse and longitudinal coordinates respectively. At the output surface $z = L$, the *diffuse* light transitions to the free-space propagation. The intensity $I_T(y)$ transmitted through the slab as a function of the transverse coordinate $y$ can be found from the continuity of the flux found as $\mathbf{J}(\mathbf{r}) = -D\nabla \langle I(\mathbf{r}) \rangle$ inside the slab and $J(\mathbf{r}) = cI_T(\mathbf{r})$ in the free space. Applying this condition at the output surface [13], we find

$$I_T(y) = (1/c)J(\mathbf{r})|_{z=L} = -(D/c)\nabla_z \langle I(\mathbf{r}) \rangle|_{z=L}, \tag{2}$$

where $c$ denotes the speed of light. We note that the more accurate boundary conditions, which include the so-called extrapolation length, see e.g. Sec. A5.2 in Ref. [13], should only introduce corrections that are smaller by the factor $\ell/L \ll 1$.

For the cases with inclusions, we represent the solution as:

$$\langle I(\mathbf{r}) \rangle = \langle I^{(0)}(\mathbf{r}) \rangle + \langle \Delta I(\mathbf{r}) \rangle \tag{3}$$

where $\langle I^{(0)}(\mathbf{r}) \rangle$ is the solution of the diffusion equation in the pure slab and $\langle \Delta I(\mathbf{r}) \rangle$ is the correction in the intensity due to the inclusion. To obtain an analytical expression for the correction term, we use the method described in Ref. [21]. For an object located at $\mathbf{r}_c = (0, z_c)$ inside the medium, the intensity can be described by a multipole expansion, in analogy with the electrostatics [21–23], where the electric potential also satisfies the Laplace equation. For a lossless medium, the lowest order dipole term describes disturbance of the intensity far enough from the object. It can be written as

$$\langle \Delta I(\mathbf{r}) \rangle = \mathbf{p} \cdot \nabla_{\mathbf{r}_c} G^{(0)}(\mathbf{r}, \mathbf{r}_c), \tag{4}$$

where $G^{(0)}(\mathbf{r}, \mathbf{r}')$ is the Green function for the pure slab, i.e. it is the solution of $\nabla_{\mathbf{r}}^2 G^{(0)}(\mathbf{r}, \mathbf{r}') = -\delta(\mathbf{r} - \mathbf{r}')$ with zero boundary conditions at the slab boundaries. The dipole moment in Eq. (4) is induced by the incident intensity so that $\mathbf{p} = -P\nabla_{\mathbf{r}_c} \langle I^{(0)}(\mathbf{r}_c) \rangle$. The intrinsic "polarizability" for a reflecting cylindrical inclusion $P^{(I)} = 2\pi R_1^2$ can be found by applying proper boundary conditions at the cylinder surface [23]. Using same method for the core-shell structure, we find

$$P^{(II)} = 2\pi R_2^2 \times \frac{(R_2^2 + R_1^2)/(R_2^2 - R_1^2) - D'/D}{(R_2^2 + R_1^2)/(R_2^2 - R_1^2) + D'/D} \tag{5}$$

where $R_{1,2}$ are the inner and outer radii of the shell and $D'$ is the diffusion coefficient of the shell. It is implicitly assumed that the size of the inclusion is much greater than $\ell$. In general, a different boundary condition at the surface of the cylindrical inclusion would result in a modified Eq. (5), whereas the rest of our calculations remain unaffected.

Using the Langevin approach [18, 24, 25], the long-range intensity correlation $C_2(\mathbf{r}_1; \mathbf{r}_2)$ has been related to the ensemble average intensity $\langle I(\mathbf{r}) \rangle$ and the Green function $G(\mathbf{r}; \mathbf{r}')$ of the diffusion equation as

$$C_2(\mathbf{r}_1, \mathbf{r}_2) = \frac{4}{k_0\ell} \frac{\int_\Omega d\mathbf{r} \, \nabla_{\mathbf{r}} G(\mathbf{r}_1, \mathbf{r}) \cdot \nabla_{\mathbf{r}} G(\mathbf{r}_2, \mathbf{r}) \langle I(\mathbf{r}) \rangle^2}{\langle I(\mathbf{r}_1) \rangle \langle I(\mathbf{r}_2) \rangle}. \tag{6}$$

In the Langevin approach, proposed in the context of mesoscopic electron transport by Spivak and Zyuzin [26], the long-range correlations are obtained from the short-range ($\Delta r \lesssim \ell$) correlations for fluxes. The result in Eq. (6) is *due wave interference* of paths with crossings in Fig. 1b; its wave nature is seen from the $(k_0\ell)^{-1}$ factor, where $k_0 = 2\pi/\lambda$ is the wave number. As in Eq. (1), Eq. (6) neglects surface corrections [24], which are small in the limit $L \gg \ell$ considered in this work. The integration in Eq. (6) is taken over the entire volume of the system $\Omega$.

In analogy with Eqs. (3,4), the Green function in the systems with inclusions can be found as

$$G(\mathbf{r}, \mathbf{r}') = G^{(0)}(\mathbf{r}, \mathbf{r}') - P \nabla_{\mathbf{r}_c} G^{(0)}(\mathbf{r}_c, \mathbf{r}') \cdot \nabla_{\mathbf{r}_c} G^{(0)}(\mathbf{r}, \mathbf{r}_c). \tag{7}$$

In what follows, we use Eqs. (3,6) to compute the diffuse intensity and the long-range correlation for the three cases of interest.

For the pure slab illuminated by a plane wave, the solution of the diffusion equation is a linear function

$$\langle I^{(0)}(\mathbf{r}) \rangle = \langle I^{(I)}(z, L) \rangle = I_0(1 - z/L). \tag{8}$$

The Green function can be easily found by e.g. Fourier transformation over the transverse variable [13]

$$G^{(0)}(\mathbf{r}, \mathbf{r}') = -\int_{-\infty}^{\infty} \frac{\sinh[kz_<]\sinh[k(L - z_>]}{k\sinh[kL]} e^{ik(y-y')} \frac{dk}{2\pi}, \tag{9}$$

where $z_< = \min[z, z']$, and $z_> = \max[z, z']$. The diffuse intensity in the presence of a reflecting cylinder at $\mathbf{r}_c = (z_c, y_c \equiv 0)$ can be found by substituting Eqs. (8,9) into Eqs. (4)

$$\langle \Delta I^{(I)}(\mathbf{r}) \rangle = -\frac{P^{(I)} I_0}{4L^2} \left( \frac{\sin[\pi(z - z_c)/L]}{\cosh[\pi y/L] - \cos[\pi(z - z_c)/L]} + \frac{\sin[\pi(z - (L - z_c))/L]}{\cosh[\pi y/L] - \cos[\pi(z + z_c)/L]} \right). \tag{10}$$

This negative correction scales as a ratio between the area $\pi R_1^2$ of the inclusion and square of the slab thickness. It leads to a shadow directly behind the cylinder that was studied experimentally and theoretically in Ref. [21] where Eq. (10) was also obtained.

To cloak a cylindrical inclusion, Ref. [4] proposed to modify the diffusion coefficient in the surrounding cylindrical shell as

$$D'/D = (R_2^2 + R_1^2)/(R_2^2 - R_1^2). \tag{11}$$

Under this cloaking condition, the polarizability in Eq. (5) vanishes and our method confirms that indeed

$$\langle \Delta I^{(II)}(\mathbf{r}) \rangle = 0. \tag{12}$$

For a slab, the long-range correlation is found by substituting Eqs. (8,9) into Eq. (6). For observation points at the output surface, a compact expression has been obtained [18]

$$C_2^{(0)}(y_1, L; y_2, L) = \frac{1}{k_0 \ell} \int\limits_{-\infty}^{+\infty} \frac{dk}{2\pi} \frac{-2kL + \sinh(2kL)}{k \sinh^2(kL)} \cos[k(y_1 - y_2)], \tag{13}$$

which we use as a reference. The presence of a reflecting cylinder can be treated perturbatively, it introduces corrections of two types

$$\Delta C_2^{(I)}(\mathbf{r}_1, \mathbf{r}_2) \simeq \frac{1}{k_0 \ell} \frac{P^{(I)}}{L^2} F(\mathbf{r}_1, \mathbf{r}_2) + \Delta C_2^{EV}(\mathbf{r}_1, \mathbf{r}_2). \tag{14}$$

The first term is the leading term due to $\langle \Delta I(\mathbf{r}) \rangle$ and $\Delta G(\mathbf{r}, \mathbf{r}')$ in Eqs. (4,7) respectively, where $P^{(I)}/L^2$ is a small parameter and $F(\mathbf{r}_1, \mathbf{r}_2)$ is a function with the amplitude on the order of unity. The second contribution in Eq. (14) originates from the reduction of the integration area (an exclusion volume) $\Omega \rightarrow \Omega - \Omega_c$ in Eq. (6). This negative contribution is proportional to the area of the cylinder $\Omega_c$ and can be estimated as $\Delta C_2^{EV}(\mathbf{r}_1, \mathbf{r}_2) \propto (1/k_0 \ell) \times (\pi R_1^2/L^2)$. Therefore, although two terms in Eq. (14) have different origin, they have comparable amplitudes. The analytical computation of $\Delta C_2^{(I)}(\mathbf{r}_1, \mathbf{r}_2)$ leads to a cumbersome expression, so we defer the further analysis of this case to the numerical section below.

Surrounding a cylindrical object with a cloak negates its "polarizability" $P^{(I)} \rightarrow P^{(II)} = 0$ when the condition in Eq. (11) is met. As a result, the first term in Eq. (14) that is proportional to $P^{(II)}$ vanishes, whereas the exclusion volume in the second term is increased to also include the cloak. Assuming that the integrand in Eq. (13) varies slowly around $\mathbf{r}_c$, we obtain the following analytical result for observation points at the output surface of the slab

$$\Delta C_2^{(II)}(y_1, L; y_2, L) = -\frac{1}{k_0 \ell} \frac{\pi R_2^2}{L^2} \frac{\pi^2 (L - z_c)^2}{4 \pi L^2} \times \\ \mathrm{Re} \left[ \cos^{-2} \frac{\pi (iy_1 + z_c)}{2L} \cos^{-2} \frac{\pi (iy_2 - z_c)}{2L} \right] \tag{15}$$

When $y_1 = y_2$, $\Delta C_2^{(II)}(y, L; y, L)$ represents a correction to the fluctuations of intensity at the output surface. We find that the transverse coordinate of the cloaked object corresponds to the maximum correction ($y = y_c \equiv 0$ in our case). The depth and the size of the object can be determined from the absolute value $|\Delta C_2^{(II)}(0, L; 0, L)| \equiv \delta$ and the full width at half maximum $\Delta y_{FWHM}$ of the correction: $z_c/L = (2/\pi) \cos^{-1}[(\sqrt{2} - 1)^{1/2} \sinh(\pi \Delta y_{FWHM}/L)]$ and $\pi R_2^2/L^2 = \delta 4 k_0 \ell L^2/(L - z_c)^2 \cos^4(\pi z_c/2L)$. Table 1 summarizes the results of our calculations. It shows that the cloak removes the correction in the transmitted intensity profile but not in the long-range correlations. Moreover, the magnitude and the scaling of the correction is the same with and without the cloak.

To confirm the analytical results derived above, we obtained the intensity $\langle I(\mathbf{r}) \rangle$ and the long-range intensity correlation $C_2(\mathbf{r}, \mathbf{r})$ numerically. First, we used Comsol Mutiphysics [19] to compute intensity $\langle I(\mathbf{r}) \rangle$ and the Green function $G(\mathbf{r}, \mathbf{r}')$ as solutions of the homogeneous diffusion equation and the diffusion equation with a point source respectively. Secondly, we used Eq. (6) to calculate the correlation.

| | Slab | | Cylinder (I) | Cloak (II) |
|---|---|---|---|---|
| $\langle I^{(0)} \rangle$ | $\propto \frac{\ell}{L}$ | $\langle \Delta I \rangle$ | $\propto \frac{\ell}{L} \times \frac{\pi R_1^2}{L^2}$ | $0$ |
| $C_2^{(0)}$ | $\propto \frac{1}{k_0 \ell}$ | $\Delta C_2$ | $\propto \frac{1}{k_0 \ell} \times \frac{\pi R_1^2}{L^2}$ | $\propto -\frac{1}{k_0 \ell} \times \frac{\pi R_2^2}{L^2}$ |

**Table 1.** Summary of the analytical results. The long-range intensity correlation contains the size and location (see Eq. (15)) information of the cloaked object.

Figures 2(a-c) show the numerically-computed intensity $\langle I(\mathbf{r}) \rangle$ and the corresponding transverse profiles at the output surface $\langle I(y, L) \rangle$, Fig. 2(d), for: slab geometry (red lines); the slab geometry with an embedded reflecting cylinder (black lines); and the slab geometry with an embedded cloaked cylinder (blue lines). In the latter two cases, the inclusions were placed in the middle of the slab. For the results in Fig. 2(a-c,e-g), the thickness was $L/\ell = 20.4$. The radii of the cylinder and cloak were chosen as $R_1/\ell = 4.5$ and $R_2/\ell = 6.8$, such that their ratio is $R_2/R_1 = 1.5$. The diffusion coefficient of the cloak satisfied Eq. (11).

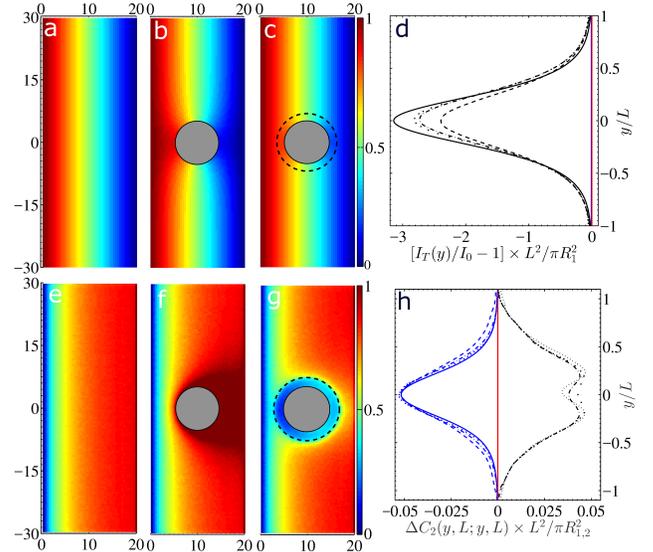

**Fig. 2.** Panels (a-c) show the diffuse $\langle I(\mathbf{r}) \rangle$ intensity computed by solving the diffusion Eq. (1) numerically for the slab, bare reflecting cylinder and the cloaked cylinder respectively. Distances are given in units of $\ell$. (d) shows the the transmitted intensity $I_T(y)$ profile at the output surface, c.f. Eq. (2). The result for the pure slab, $I_T^{(0)}(y)$, is used for normalization to demonstrate the agreement with the theoretical prediction in Eq. (10). The red, black and blue curves correspond to the three geometries shown in (a-c) respectively. The dashed, dot-dashed and dotted black curves correspond to slabs with $L/\ell = 20$, $34$, $41$; solid line is the analytical result in Eq. (10). Panels (e-g) show the amplitude of the long-range intensity correlation $C_2(\mathbf{r}, \mathbf{r})$ computed numerically from Eq. (6) and normalized to the maximum value in the case of the pure slab. Panel (h) shows $\Delta C_2(L, y; L, y)$ using the same color scheme as in (d); solid blue line is the analytical result from Eq. (15).

As predicted analytically by Eq. (10), there exists a diffusive shadow behind the cylinder, c.f. Figs. 2(b,d), where the intensity is reduced. In contrast, the cloaked cylinder has no shadow, concealing the presence of the cylinder in accordance with Eq. (12). To verify the theoretical scaling in Eq. (10), solid black line in Fig. 2(d), we normalized the numerical intensity profiles by $\pi R_1^2/L^2$. The dashed, dot-dashed, dotted black lines are the normalized intensities at the output surface in the slabs with an embedded cylinder and $L/\ell = 20$, 34 and 41, respectively. The numerical results approach the analytical solution with an increase of $L$, when the dipole approximation used in deriving Eq. (10) is asymptotically satisfied.

In general, the long-range correlation $C_2(\mathbf{r}_1, \mathbf{r}_2)$ depends on two spatial coordinates $\mathbf{r}_1$ and $\mathbf{r}_2$. Figures 2(e-g) depict the diagonal part $C_2(\mathbf{r}, \mathbf{r})$ computed numerically using Eq. (6). This quantity corresponds to the non-Rayleigh contribution to the intensity fluctuation. In the slab geometry, Fig. 2(e), $C_2(\mathbf{r}, \mathbf{r})$ depends only on $z$-coordinate as $C_2(z;z) \propto (z/L)(1 - 2z/3L)$ [27].

In the case of slab with a cylinder, Fig. 2(f), the long-range correlation is enhanced. This is because photon path lengths increase in order to circumvent the cylinder and, thus, the probability of path crossings, responsible for the correlation, is also increased. The amplitude of the correction $\Delta C_2(L, y; L, y)$, computed for slabs of different thickness (black lines in Fig. 2(h)), scales as $\propto (k_0 \ell)^{-1} \times \pi R_1^2/L^2$, as predicted in Table 1. This sets the level of precision for observation of the discussed effect.

The diminished diffusion coefficient of the cloak, Eq. (11), reduces the intensity correlation below the reference level of the pure slab, see Fig. 2(g). This agrees both qualitatively and quantitatively with the prediction of Eq. (12). Indeed, as shown in Fig. 2(h), our analytical result describes $\Delta C_2(L, y; L, y)$ computed for slabs of different thickness. Both the profile and the size scaling of the correction term are in agreement.

In the realistic systems, an absorption tends to suppress the transmission and its fluctuations. However, the correlation in Eq. (6) is normalized by the average intensity. This largely compensates the exponential attenuation factor associated with absorption. Hence, the long-range correlations persists at the distances longer than the characteristic absorption length [28], and we expect that the effect predicted in our work will depend only weakly on absorption losses. The detailed account of loss goes beyond the scope of the current work.

Analogy between the diffusion and the electrostatics problems offers insights into the nature of the diffusive cloak. It shows that the intensity perturbation due to embedded objects can be suppressed by cancellation of "multipoles". In case of cylindrical (spherical in 3D) inclusions, a cancellation of the dominant dipole term leads to cloaking, Eq. (4). Such an approach is also analogous to ballistic cloaks employing metamaterials [29].

In this work, we exploit the fact that diffusion is only an approximate description of the electromagnetic wave transport in turbid media. The intrinsic wave effects are responsible for such phenomena as speckles, the intensity fluctuation and correlation. In particular, the non-local nature of the long-range correlation offers an opportunity to detect the diffusive cloak. Crossings of the diffusive paths and the subsequent interference serve as "beacons", which give rise to the intensity correlation detectable at the output surface of the medium. In Ref. [30], ultrasound focus was used to shift the wavelength of light and, thus, to temporarily create a source inside a strongly scattering sample. Such a technique can offer an alternative approach to detecting the diffusive cloak. However, unlike the long-range correlation employed in our work, scanning ultrasonic focus would be a serial process. Lastly, an entirely different kind of a "beacon" – Cherenkov radiation produced by the fast-moving charged particles – was proposed for detection of the ballistic cloak based on transformation optics [31].

Aside of detection of diffusive cloak, our work also offers an intriguing possibility to manipulate the second-order statistics (correlations/fluctuations) independently from the first-order statistics (average intensity). Finally, the results can be trivially generalized to 3D systems and for waves of different nature – acoustic, electronic, seismic, etc.

## FUNDING.

This work has been supported by National Science Foundation grant DMR-1205223.

## ACKNOWLEDGMENT.

The authors thank Hui Cao and Raktim Sarma for several insightful discussions.